\documentclass[aps,prb,reprint,superscriptaddress,showpacs, floatfix]{revtex4-1}
\usepackage{graphicx}
\usepackage{color}
\usepackage{transparent}
\usepackage{hyperref}
\usepackage{natbib}
\usepackage{dcolumn}
\newcolumntype{d}[1]{D{.}{.}{#1}}
\begin{document}
\title{The Ultrafast Optical Response of the Amorphous and Crystalline States of the Phase Change Material Ge$_2$Sb$_2$Te$_5$}
\author{T. A. Miller}
\affiliation{ICFO-Institut de Ciencies Fot\`oniques, The Barcelona Institute of Science and Technology, 08860 Castelldefels (Barcelona), Spain}
\author{M. Rud\'e}
\affiliation{ICFO-Institut de Ciencies Fot\`oniques, The Barcelona Institute of Science and Technology, 08860 Castelldefels (Barcelona), Spain}
\author{V. Pruneri}
\affiliation{ICFO-Institut de Ciencies Fot\`oniques, The Barcelona Institute of Science and Technology, 08860 Castelldefels (Barcelona), Spain}
\affiliation{ICREA - Instituci\'{o} Catalana de Recerca i Estudi Avan\c{c}ats, Barcelona, Spain}
\author{S. Wall}
\email[Corresponding author:]{simon.wall@icfo.es}
\affiliation{ICFO-Institut de Ciencies Fot\`oniques, The Barcelona Institute of Science and Technology, 08860 Castelldefels (Barcelona), Spain}

\date{\today}

\begin{abstract}
We examine the ultrafast optical response of the crystalline and amorphous phases of the phase change material Ge$_2$Sb$_2$Te$_5$ below the phase transformation threshold. Simultaneous measurement of the transmissivity and reflectivity of thin film samples yields the time-dependent evolution of the dielectric function for both phases. We then identify how lattice motion and electronic excitation manifest in the dielectric response. The dielectric response of both phases is large but markedly different. At 800 nm, the changes in amorphous GST are well described by the Drude response of the generated photo-carriers, whereas the crystalline phase is better described by the depopulation of resonant bonds. We find that the generated coherent phonons have a greater influence in the amorphous phase than the crystalline phase. Furthermore, coherent phonons do not influence resonant bonding. For fluences up to 50\% of the transformation threshold, the structure does not exhibit bond softening in either phase, enabling large changes of the optical properties without structural modification.
\end{abstract}

\maketitle

\section{Introduction}
Phase change materials (PCMs) based on alloys of Ge, Sb and Te can be easily, rapidly, and reversibly switched between ordered-crystalline and disordered-amorphous crystallographic arrangements. Switching between states is typically achieved by a heat-quench cycle, which can be laser-~\cite{Weidenhof2001, Liu2012b} or current-induced~\cite{Lankhorst2005a, Loke2012b}. Amorphization is achieved by heating the crystalline phase above the melting point followed by rapid cooling. Re-crystallization is achieved by heating above the glass transition temperature and then slowly cooling~\cite{Wuttig2007a}.

The crystallographic switching process is accompanied by an unusually large change in the optical and electronic properties of the material and makes PCMs important for many applications, such as optical data storage~\cite{Wuttig2007a}, non-volatile memories~\cite{Wuttig2005} and photonics~\cite{Rude2013, Rios2014}. The large optical contrast is attributed to the so-called resonant bonding in the crystalline phase~\cite{Lucovsky1973, Shportko2008a}. The alignment of $p$-orbitals over next-nearest neighbors enhances the optical matrix elements, which, in turn, dramatically enhances the optical properties~\cite{Huang2010a}. The lack of medium-range $p$-orbital alignment defines the amorphous phase, which otherwise has similar local atomic arrangements and bonding. Within this framework, the effect of the phase transition on the dielectric function of the prototypical PCM, Ge$_2$Sb$_2$Te$_5$ (GST) is known. The peak in optical absorption of the crystalline state resulting from bonding-antibonding transitions~\cite{Shportko2008a} induces a large real part of the dielectric function in the low-frequency limit~\cite{Huang2010a}. Upon amorphization, the peak in absorption shifts to higher energies and is reduced.  This corresponds to a reduction of the real part of the dielectric function when viewed at a single low frequency below the bonding-antibonding transition.

In equilibrium, the optical properties are linked to the structural transition. However, out of equilibrium this does not have to be the case. Recent measurements of the time-resolved optical and structural changes in GST demonstrated that the some of the desirable optical properties of the crystalline phase change materials can be switched faster than the time required for structural amorphization~\cite{Waldecker2015b}.

Switching the optical properties without structural motion would present new opportunities for PCMs. The PCM device lifetimes are limited by ionic migration and accumulated stress acquired during the structural transition. Secondly, structural motion is inherently slow, and removing the melting/crystallization cycle could dramatically improve device speed, at the cost of a permanent change. Therefore, large ultrafast changes in the optical properties will be important for making high speed optical modulators~\cite{Rude2015} and other such photonic devices. Furthermore, rapid optical modification without structural switching from one of two stable starting points may open the possibility of novel device design. One may then choose the starting phase depending on the application-specific optimal dielectric response.

In this paper we examine and compare, in detail, the changes in optical properties of amorphous {\em and} crystalline phases of GST following excitation with femtosecond laser light when permanent switching does {\em not} occur. We consider an excitation regime where large optical changes are possible without permanent atomic rearrangement, and we consider both starting crystallographic states. We deduce the influence of free photoexcited carriers and structural changes on the materials' optical properties. We find that the changes observed in the amorphous phase are consistent with photo-generation of free carriers in a semiconductor, whereas the response of the crystalline state is better described by breaking of resonant bonds. Furthermore, just as the intrinsic vacancies of GST play a major role in the static optical and electrical properties~\cite{Wuttig2007b}, they are also play a key role in the optically-induced structural dynamics. 

\section{Static Optical Properties}

\begin{figure}
\centering
\includegraphics{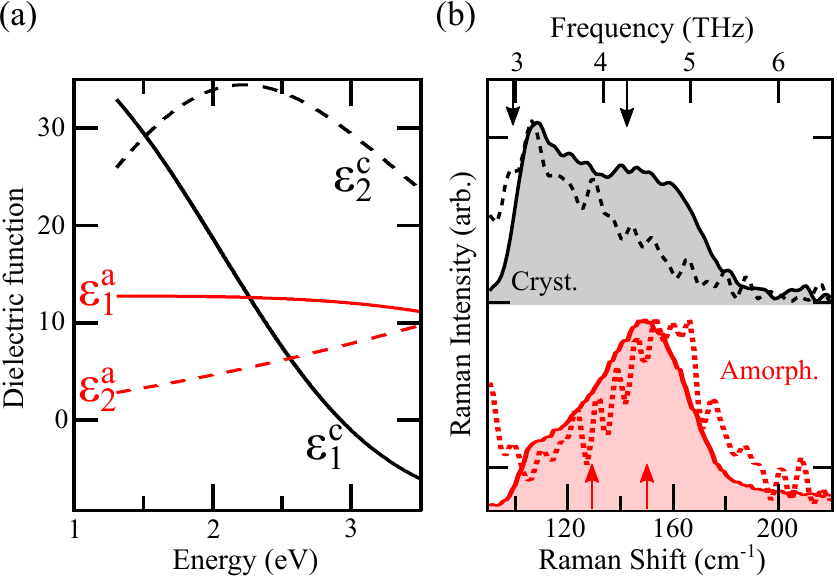}
\caption{(color online) (a) Ellipsometry measurements of the dielectric function for the crystalline (black) and amorphous (red) phases: $\epsilon^{c/a} = \epsilon^{c/a}_1 + i\epsilon^{c/a}_2$. (b) Raman measurements using a 785\,nm laser are shown with solid lines. Notice the increased weight at low frequencies for the crystalline phase. Arrows correspond to phonon frequencies used in Eq.~\ref{eq:p}. The Fourier transforms of the phonons observed at the highest excitation in this work are shown with dashed lines and agree well with the static Raman measurements.\label{fig:Static}}
\end{figure}

30\,nm thick samples of amorphous Ge$_2$Sb$_2$Te$_5$ (GSTa) were deposited on fused silica substrates by RF co-sputtering and capped with 10\,nm of Si$_3$N$_4$. The optical and Raman spectra are shown in Figure~\ref{fig:Static}. The exact optical properties of the sample are quite sensitive to sample preparation, and the values we observe are consistent with the spread of values reported in the literature~\cite{Lee2005,Shportko2008a, Park2009}.  The GSTa Raman spectrum shows an asymmetric peak around 4.5\,THz with more weight on the high energy side. The as-deposited amorphous phase has slightly different optical and structural properties than the melt-quenched amorphous state (detailed comparison in supplementary information of ref~\onlinecite{Waldecker2015b}). The crystalline state (GSTc) can be achieved by annealing the as-deposited sample on a hot plate at 200\,$^\circ$C for 1\,h, heating at 10\,$^\circ$C/min. The phase change, in addition to drastically changing the dielectric function, also shifts the peak of the Raman/phonon spectrum to lower energies and further transfers spectral weight to the low energy side of the peak. 

We note that the Raman spectrum of the crystalline phase was also sensitive to the heating and cooling rate from the as-deposited amorphous phase. A crystalline phase with more pronounced peaks could also be formed. However, we chose a sample with the broad phonon/Raman spectrum because it matches that obtained from cw-laser crystallization~\cite{Kolobov2004} and is thus the more technologically relevant sample. 

\section{Transient Experimental Setup}
Pump probe measurements were performed in the reversible regime with 40\,fs laser pulses at a central wavelength of 800\,nm. The pump spot was focused to 230\,$\mu$m and the probe to 20\,$\mu$m FWHM to ensure that a uniformly excited sample volume was probed. The beams were measured on a CCD camera, and the fluence reported here is that obtained by the weighted average of the pump that the probe experiences. Pump and probe beams were separated by a small angle and impinged on the sample at near-normal incidence. Pump and probe were cross-polarized, and polarizers were used to filter out residual pump scatter from the detectors. The use of thin samples on a transparent substrate allowed us to measure both changes in the reflectivity ($R$) and transmissivity ($T$), which were acquired simultaneously for each delay. No baseline subtraction was applied to the data, as baseline stability was continuously monitored for any sign of sample change. As an additional check, we also ensured that the low fluence transients were unchanged after completing measurements at high fluence.

The measurements were performed at a 40\,Hz repetition rate, substantially lower than many pump-probe measurements, to ensure that the sample fully recovered between laser pulses, including dissipation of any thermal accumulation. Consequently, no changes in material properties were observed after several hours of excitation for the fluences reported here. At higher repetition rates and higher powers, gradual laser-induced changes to the coherent phonon spectrum of both samples were observed. These changes were similar to, but still distinctly different from the thermally induced effects on the static Raman spectrum. The changes were consistent with the emergence of the modes of elemental Te, indicating that segregation had occurred. Segregation has been observed in other Te-based compounds~\cite{Shimada2014}. An example of such laser-induced changes is shown in Figure~\ref{fig:phononSoftening} and is discussed in more detail in Section~\ref{sec:dis}.

\section{Measured Results}

\begin{figure}
\centering
\includegraphics{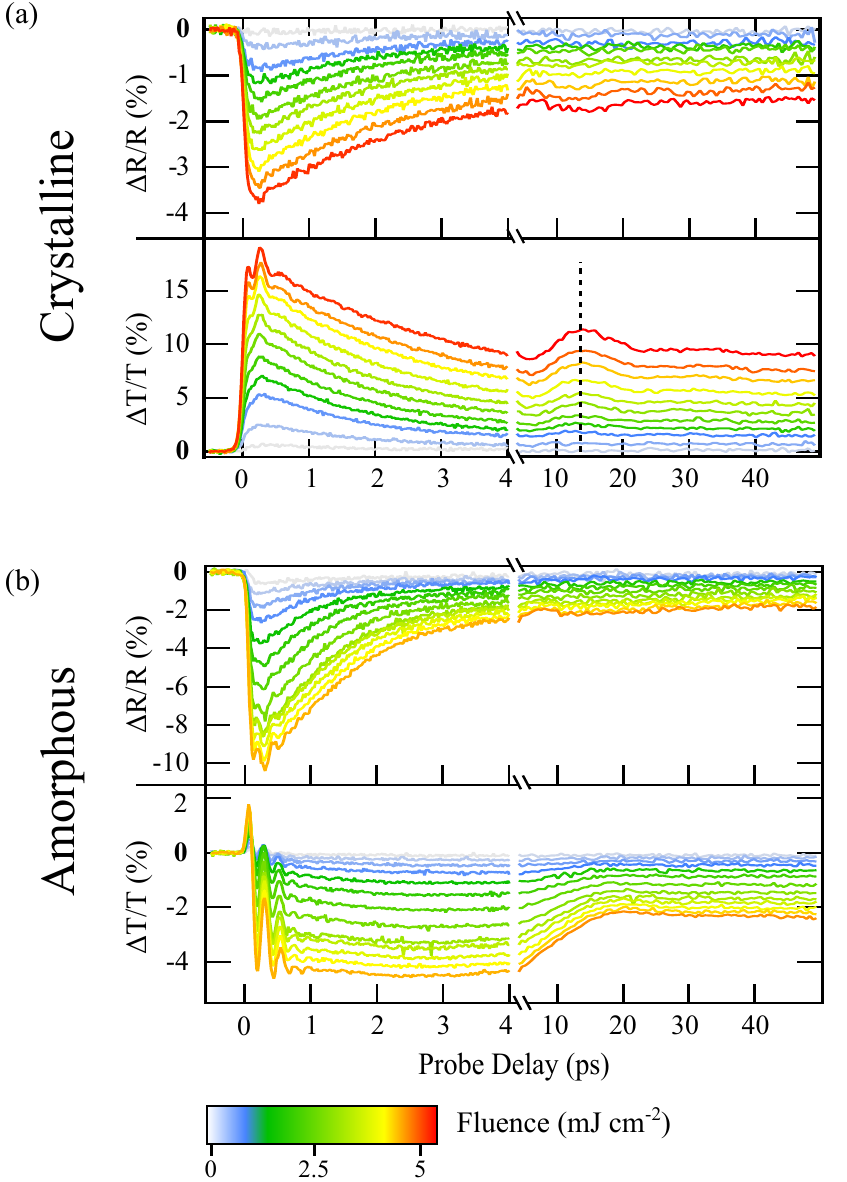}
\caption{(color online) Measured changes in reflectivity and transmissivity for (a) crystalline and (b) as-deposited amorphous states of GST. The amplitude of the change is linear in pump fluence. Note that no softening of the acoustic phonon (marked with dashed line) is observed.\label{fig:RTgstAC}}
\end{figure}

The time resolved changes in $R$ and $T$ for GSTc and GSTa are shown in Figure~\ref{fig:RTgstAC} for several pump fluences well below the threshold for permanent switching of the material properties (\textless14\,mJ\,cm$^{-2}$, ref~\onlinecite{Waldecker2015b}). The two phases show very different dynamics. On short timescales, the crystalline phase shows a large increase in transmission (up to 20\%) and a smaller decrease in reflection (-4\%) . A weak coherent phonon can be observed in transmission, but is barely visible in the reflected signal. At longer times ($\approx$12\,ps), a secondary peak is observed both in transmission and more weakly in the reflection. This peak corresponds to a acoustic breathing mode of the film,  found to be 44$\pm2$\,GHz. Given the known film thickness, we calculate the sound velocity to be 2.6 $\pm0.3$\ nm/ps in GSTc, lower than the literature value of 3.16\,nm/ps~\cite{Lyeo2006}. Furthermore, the temporal position of the acoustic phonon peak did not change with pump fluence, as indicated by the dashed line in Figure~\ref{fig:RTgstAC}a. Therefore, in the regime measured here, the sound velocity is constant and the lattice remains rigid.

In contrast to GSTc, the amount of light statically reflected and transmitted both decrease in GSTa. The photo-induced fractional changes are largest in the reflected signal, which show decreases as large as 10\%. The transmission changes are smaller at 4\%. The coherent phonon signal is strongest when measured in transmission, where it dominates the signal at short times, but it is now also observable in the reflected signal. At longer timescales the acoustic phonon is no longer observed, which is a consequence of the increased damping rate and slower speed of sound for acoustic waves in the amorphous sample~\cite{Lyeo2006}. Furthermore, the dynamics observed in transmission seem more complex. The change is initially positive before rapidly becoming negative within a few tens of femtoseconds, with only small changes in the subsequent 3\,ps of evolution.

\begin{figure}
\centering
\includegraphics{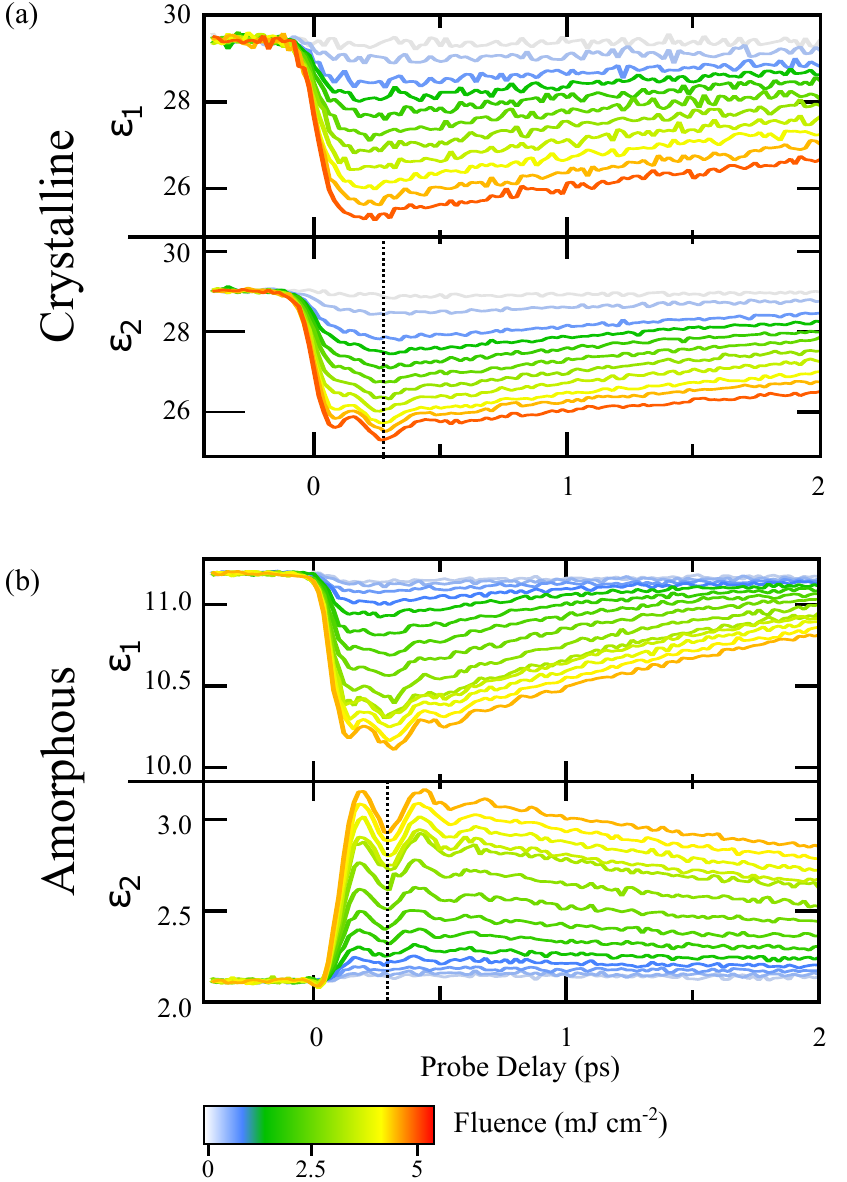}
\caption{(color online) The time-dependent dielectric function of crystalline (a) and amorphous (b) GST at short times. In crystalline GST the optical phonon modifies only $\epsilon_2$, but modifies both terms in amorphous GST. Note the large magnitude of the effect for crystalline GST. The optical phonons (dashed lines) shows no frequency shift in either phase.\label{fig:dielectric}}
\end{figure}

Although the reflectivity and transmissivity reveal many salient features of the dynamics, interpreting the results is complicated by the fact that these quantities are not fundamental material properties. Therefore, we use these data to obtain the complex dielectric function by the transfer matrix method as done previously~\cite{Waldecker2015b}. The optical properties of the Si$_3$N$_4$ layer and SiO$_2$ substrate were held constant at literature values, and only the dielectric properties of the GST layer were allowed to change. We also assumed that the thickness of the GST layer did not change. As this is not strictly true due to the presence of the acoustic phonon, we limit the evaluation of the dielectric function to the first few picoseconds where the thermal expansion remains small.

The extracted time-dependent dielectric functions are shown in Figure~\ref{fig:dielectric}. In the crystalline phase, both the real ($\epsilon_1$) and imaginary ($\epsilon_2$) parts of the dielectric function show large decreases with photoexcitation, and the coherent phonon is only observed to modulate the imaginary part. In the amorphous phase, the real part of the dielectric function decreases whereas the imaginary part increases. Furthermore, the coherent phonon is observed on both the real and imaginary terms. It is interesting to note that the complex dynamics observed in the transmission signal of GSTa (Figure~\ref{fig:RTgstAC}) are simplified when converted to the dielectric function. After the initial phonon oscillations, the transmission remains almost constant for several picoseconds. This could lead to the conclusion that the system is not evolving; however, conversion to the dielectric function makes clear that the system continues to respond with an almost equal-but-opposite response in both components of the dielectric function.

\section{Modeling the Optical Response}
Theoretical modeling of the optical properties in non-equilibrium systems is a rapidly developing field. The effects of the photo-excited electron-hole plasma~\cite{Benedict2001} and phonon displacements~\cite{Katsuki2013} on a material's dielectric function can be accurately calculated in specific systems. However, the time evolution of these properties still remains out of reach. Furthermore, these state-of-the-art techniques are difficult to apply to GST, because a large unit cell is needed to capture the amorphous phase making it computationally too expensive.

\begin{figure}
\centering
\includegraphics{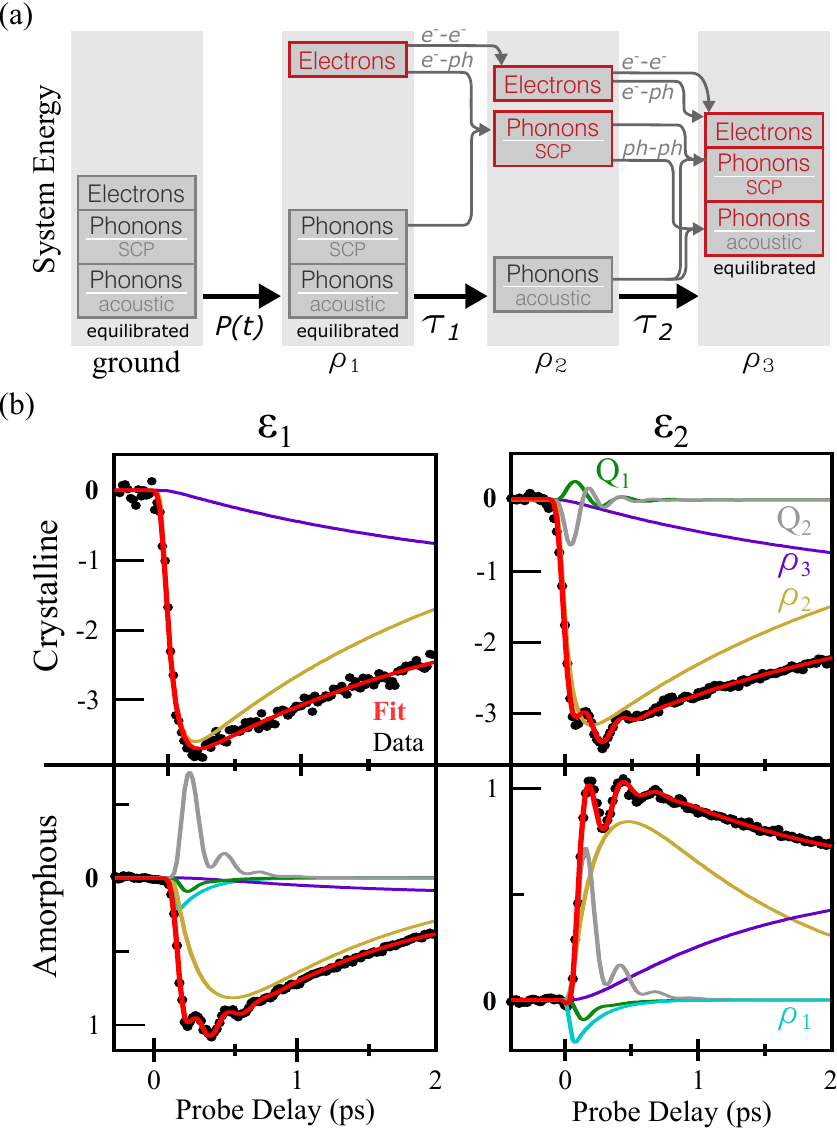}
\caption{(color online) Schematic representation of the model (a). The system is electronically excited from the ground state by the pump laser ($\rho_1$). That energy first couples to coherent phonons (SCPs) ($\rho_2$), and later couples to the lattice at large ($\rho_3$) Many energy pathways (e.g. electron-electron scattering, electron-phonon scattering, etc., grey lines) are represented with a single time constant. A representative fit using Eqs.~\ref{eq:ee} and~\ref{eq:ep} to the time-dependent dielectric function of GSTc and GSTa (b). The temporal evolution of the individual fit components is also shown. The fluence is 4.6\,mJ\,cm$^{-2}$.\label{fig:dEPgstA}}
\end{figure}

Instead, we construct a phenomenological model. The purpose of our model is to extract the amplitudes and timescales of the various processes that modify the dielectric function. Additionally, this model is chosen to fit both phases of GST to permit comparison, and the model is able to accurately reflect time-zero dynamics to correctly capture the phonon frequencies and any potential softening. To that end we construct a model which separates the energy flow within the material into three sequential reservoirs, $\rho_\mathrm{1}, \rho_\mathrm{2}, \rho_\mathrm{3}$ (Figure~\ref{fig:dEPgstA}). The different reservoirs can be loosely assigned to different underlying physical states. $\rho_\mathrm{1}$ represents the energy in the initially excited electron-hole pairs, which is supplied by the pump laser, $P(t)$. The photoexcited carriers thermalize through scattering amongst themselves and with a subset of strongly coupled phonons (SCP), represented by $\rho_\mathrm{2}$. Although the scattering time for the electrons and holes is likely different, we represent them with an average time constant $\tau_1$. Finally the carriers and SCPs scatter with the other acoustic and optical phonons to thermalize the system, represented as state $\rho_\mathrm{3}$. Again, although there are multiple scattering processes involved, we represent the overall process by a second time constant $\tau_2$. The sequential reservoirs give rise to the following set of coupled equations:
\begin{eqnarray*} 
	\dot{\rho_\mathrm{1}} & = & P(t) - \rho_\mathrm{1}/\tau_1 \\
	\dot{\rho_\mathrm{2}} & = & \rho_\mathrm{1}/\tau_1 - \rho_\mathrm{2}/\tau_2 \\
	\dot{\rho_\mathrm{3}} & = & \rho_\mathrm{3}/\tau_2
\end{eqnarray*}

While the model is similar to a three temperature model, we stress that these rates do not directly represent thermalization rates. For example, it is known that the lattice thermalization time in the crystalline state is 2.2\,ps regardless of fluence~\cite{Waldecker2015b}. Here, lattice thermalization most closely corresponds to $\tau_2$. However, in our model this process is also averaged with other effects, such as electron-hole recombination. 

To model the optical response of electronic origin we consider the simplest case of linear coupling. This is justified as long as there is linear dependence of the transient response on fluence (i.e. below the phase transformation threshold). As a result, a set of complex coefficients $a_i$  link the reservoirs to the dielectric function, yielding the following expression:
\begin{equation}\label{eq:ee}
\epsilon_e(t) = \sum_{i=1}^3 a_i\rho_i(t),
\end{equation}
where $\rho_i$ is normalized. In principle, the $a$ coefficients depend on the probe wavelength.  

We now consider structure-based modifications of the dielectric function. The coherent lattice response is found by solving the differential equation
\begin{equation}
\ddot{Q_i} + 2\gamma_i\omega_{0i}\dot{Q_i} + \omega_{0i}^2Q_i \propto  F(t),
\label{eq:p}
\end{equation}
where $Q_i$ is the displacement of the phonon mode, and $\gamma_i$ and $\omega_{0i}$ are the damping ratio and angular frequency respectively. $F(t)$ is the force which drives the displacements, which we assume is proportional to the initially created electron-hole distribution. The phonon modes then couple to the dielectric function with another set of complex coefficients, yielding:
\begin{equation}\label{eq:ep}
\epsilon_p(t) =\sum_i b_iQ_i,
\end{equation}
where, again, the $b_i$ coefficients depend on probe wavelength. 

The total dynamic contribution of both electronic and structural effects on the dielectric function is the the sum of Equations \ref{eq:ee} and \ref{eq:ep}. The sum was fit simultaneously to both the real and the imaginary part of the dielectric function, and the result was convolved with the pump pulse temporal profile to account for finite experimental time resolution. Excellent fits can be obtained at all fluences for both material phases. Example fits are shown in Figure~\ref{fig:dEPgstA}b, as well as the temporal evolution of all of the fit components.

\section{Model Results}
The lattice oscillation of both the amorphous and crystalline phases is well described by two harmonic phonons, despite the fact that the Raman spectra shown in Figure~\ref{fig:Static}b are clearly asymmetrically broadened. Initially the phonon frequencies, damping rates, and the phonon force time constant, $\tau_1$, were allowed to vary as a function of fluence, as phonon softening has been observed in some samples of GST and related compounds~\cite{Kim2013,Hase2003b,Fons2015b,Makino2011}. However, the phonon peak position did not soften in the raw data in Figure~\ref{fig:dielectric}, and neither was a frequency shift obtained from the fit. The averages of the fitted values, summarized in Table~\ref{tab:fitps}, were then taken and a second fit iteration was performed with these parameters held constant.  

The resulting phonon frequencies are in good agreement with those observed in the static Raman spectra, marked with arrows in Figure~\ref{fig:Static}b. Also shown in Figure~\ref{fig:Static}b is the phonon spectrum of both phases for the highest level excitation. Again, excellent agreement is found between the static Raman spectra and the time-resolved data, further indicating that no significant mode softening is observed. 

\begin{table}
    \centering
    \begin{tabular}{|l|d{2.2}|d{2.2}|} 
    \hline
                        & \multicolumn{1}{c|}{ Amorphous }	& \multicolumn{1}{c|}{ Crystalline } \\
    \hline
    $\omega_1$  (THz)	& 3.9	& 2.98 \\
    $\gamma_1$ 	        & 0.26	& 0.24 \\
    \hline
    $\omega_2$ 	(THz)	& 4.5 	& 4.29 \\
    $\gamma_2$          & 0.43  & 0.24 \\
    \hline
    $\tau_1 $ (fs)	    & \multicolumn{1}{c|}{197}   & \multicolumn{1}{c|}{80} \\
    \hline
\end{tabular}

\caption{Summary of the phonon fit parameters that are independent of fluence: the phonon frequencies ($\omega$), damping constant ($\gamma$), and force constant ($\tau$). The phonon frequencies are marked with arrows in Figure~\ref{fig:Static}b. \label{tab:fitps}}
\end{table}

The time constant $\tau_1$ strongly depends on the starting phase of the material. Scattering of photo-excited carriers occurs almost twice as quickly in the crystalline phase than in the amorphous phase. As these carriers are those that create a displacive force on the lattice, a short $\tau_1$ translates into a short-lived force on the lattice.

The thermalization time, $\tau_2$, increases with fluence for both the amorphous and crystalline states. In the crystalline state, the value varies non-linearly from 1.6\,ps at low fluences to 2.2\, ps at 4\,mJ\,cm$^{-2}$ and is shown in more detail in Ref.~\onlinecite{Waldecker2015b}. In the amorphous state the recovery is faster, ranging from 0.8\,ps at low fluences to 1.2\,ps at 4 mJ\,cm$^{-2}$. As this term contains the equilibration between excited carriers and the lattice, the faster thermalization of GSTa is consistent with observations of carrier lifetime decay using THz spectroscopy~\cite{Shu2013}.

\begin{table}
 \centering
 \begin{tabular}{|c|d{3.3}d{3.3}|d{3.3}d{3.3}|}
 \hline
  & \multicolumn{2}{c|}{Amorphous} 	& 	\multicolumn{2}{c|}{Crystalline} \\
  & \multicolumn{1}{c}{\textit{Real}} 
    & \multicolumn{1}{c|}{\textit{Imaginary}} 
        & \multicolumn{1}{c}{\textit{Real}} 
            & \multicolumn{1}{c|}{\textit{Imaginary}} \\
 \hline
 $a_1/F$     & -0.07     & -0.06     &  0        &  0    \\
 $a_2/F$     & -0.26     &  0.27     & -0.75     & -0.58 \\
 $a_3/F$     & -0.02     &  0.12     & -0.29	 & -0.28 \\
 \hline
 $b_1/F$     &  0.15     &  0.17     & 0         &  0.08 \\
 $b_2/F$     & -0.22     & -0.04     & 0         & -0.13 \\
 \hline
\end{tabular}

\caption{Summary of the fit parameters that depend on fluence. The observed linear dependence is given in  units of $\Delta\epsilon$\,/mJ\,cm$^{-2}$. The $a$ terms represent the electronic response, while the $b$ terms describe lattice vibration.\label{fig:fitsummary}}

\end{table}

The remaining fit parameters describe how the electronic  and crystallographic response (the $a$ and $b$ terms) modify the dielectric function. They were found to vary roughly linearly with fluence, with small deviations at the highest fluences. The linear dependencies of the coefficients are reported in Table~\ref{fig:fitsummary}. The initial photo-excited carriers ($a_1$), have no effect in the crystalline state and a small effect in the amorphous phase. Both $a_2$ and $a_3$ modify the dielectric function in the crystalline phase more than in the amorphous phase. In the crystalline phase the changes decrease both parts of the dielectric function with increasing pump fluence, moving towards the values of the amorphous phase. This is expected for amorphization. However, the amorphous state values do not move towards those of the crystalline state, demonstrating that the initial dynamics in GSTa are not related to re-crystallization. The structural response of the material, obtained from $b$ coefficients, plays a stronger role in the modulation of the dielectric function in the amorphous phase compared to the crystalline phase, and the magnitude of the changes are similar to those of the $a$ coefficients.

\section{Discussion\label{sec:dis}}
To gain more insight into the observed processes we compare the size of the changes induced in the dielectric function with those that would be induced by the generation of free carriers in the conduction band, which is typically observed in semiconductors~\cite{Huang1998, Kim2002, Sundaram2002}. Such a response can be estimated using the Drude model. We use the scattering time and conductivity of GSTc reported in ref~\onlinecite{Mendoza-Galvan2000} for both states. Assuming that photoexciation creates 30\% of the equilibrium number of charges, for a probe wavelength of 800\, nm we obtain
\begin{eqnarray*}
    \Delta\epsilon_\mathrm{Drude,amorph} = -0.3 + 0.04i \\
    \Delta\epsilon_\mathrm{Drude,crystaline} = -2.3 + 0.30i
\end{eqnarray*}
Note that for a pure Drude response, the real term of the dielectric function is expected to decrease while the imaginary term increases for both phases of GST. This is what is seen in GSTa. Thus we conclude that $\rho_2$ is dominated by the response of the free carriers.  Also, the correspondence in the timescales of free carriers measured in the THz regime to that seen here further supports a Drude interpretation~\cite{Shu2013}. This makes the response of GSTa similar to that of amorphous GeSb~\cite{Callan2001}.

More striking is that the observed optical response of GSTc is \emph{not} Drude at 800\,nm, as the imaginary term response has the wrong polarity. In this spectral region, and unlike GSTa, GSTc optical properties are dominated by resonant bonding~\cite{Littlewood1979, Shportko2008a}. Resonant bonding is primarily responsible for the large real part of the dielectric function found in the crystalline state, shown in Figure~\ref{fig:Static}a. 

Resonant bonding can be altered two ways: by changing the carrier distribution or by changing the alignment of $p$ bonds on medium length scales~\cite{Huang2010a}. In GSTc the initial photocarriers depopulate the states responsible for the resonant bonding, inducing the large optical change. As the carriers thermalize with the lattice they further induce changes in the bond angle that temporally extends the reduction in resonant bonding. The thermal energy must dissipate before extended alignment returns and resonant bonds can be re-established. We stress that the structural reduction of resonant bonding arises from a quasi-thermal distribution of phonons with a range of wave-vectors and not the coherent optical phonons.

We now consider the effect of coherent lattice motion on the dielectric response. It is surprising that coherent optical phonons are observed in GSTc, since the NaCl-like structure (space group $Fm\bar 3 m$) with only 4$a$ and 4$b$ sites occupied~\cite{Nonaka2000} does not allow Raman-active modes. Phonons are observed because the intrinsic vacancies in GSTc break the symmetry at specific sites. This enables a range of vibrational frequencies in the Raman spectrum and is why the amorphous phase Raman spectrum is higher energy than the crystalline one, as more modes become active~\cite{Sosso2011a}. Thus, unlike the de-localized modes observed in most crystalline phases, the phonon modes in both the crystalline and amorphous states of GST should be considered localized to vacant sites. With this in mind, it is perhaps not surprising that the force time constant $\tau_1$ for the crystalline phase is significantly less than that of the amorphous phase. As the crystalline state is conductive, excited electronic states are de-localized. Thus electrons excited in the vicinity of the vacancies can quickly spread, reducing the time spent in the neighborhood of the vacancy and the time over which force is applied to the localized lattice mode. In the amorphous state electrons are more localized, which extends the time over which the lattice force is applied. Furthermore these local modes can be more easily excited in the amorphous state due to the greater presence of defects, which is a natural consequence of the increased crystallographic disorder. This also explains the larger amplitude of the $b$ terms seen in GSTa (Figure~\ref{fig:dEPgstA}b).

In addition to the force lifetime, the effect of coherent phonons on the optical properties of both phases is markedly different. In GSTc, both phonon modes modulate only the imaginary part of the dielectric function, (the $b$ terms in Table~\ref{fig:fitsummary}). To the contrary, in the amorphous phase the phonon modifies just the real part of the dielectric function (the 4.5\,THz phonon, $b_2$) or both parts equally (the 3.9\,THz phonon, $b_1$, as well as $b_2$). This difference is most likely due to the shift in the GSTc absorption resonance peak position shown in Figure~\ref{fig:Static}a. Coherent phonons typically modify optical properties through band gap or resonance modulation~\cite{Papalazarou2012, Faure2013, Leuenberger2015}. In the crystalline state, the 1.5\,eV probe photon energy is close to the absorption resonance, and thus the phonon mostly likely modifies the absorption through shifts of charge transfer resonance in this spectral region. The charge transfer resonance shifts to significantly higher energies in the amorphous phase, so absorption changes plays a smaller role in the GSTa response at the 800\,nm probe wavelength. Consequently changes in the real part of the dielectric function become more pronounced. The increased visibility of the phonon is expected in GSTa, as the Raman scattering of GSTa is also stronger.

Since resonant bonding is sensitive to bond angle distortions, it has been suggested that vibrationally exciting GSTc through laser-generated coherent phonons may be able to drive the amorphization~\cite{Kolobov2011}. De-stabilizing resonant bonding in the crystalline phase would modulate the {\em real} part of the dielectric function, which is not observed in our measurements. We believe this is because the observed phonons are due to local atomic motion near vacancies and thus may not be able to influence the angles of the more spatially-distributed resonant bonds.

When GST is excited to the transformation threshold, the lattice potential significantly changes due to melting~\cite{Mitrofanov2016}. However, both the coherent acoustic and optical phonons show no indication of softening up to approximately 50\% of single-shot fluence in the present work. This shows that the lattice potential is not significantly modified in either phase below the phase transformation threshold. This is in contrast with other measurements of similar materials in which phonon softening has been observed~\cite{Kim2013,Hase2003b,Fons2015b,Makino2011}. There  are several possible explanations for why our results differ. 

The extremely low repetition rate is used in our measurements is several orders of magnitude lower than those used in other pump probe measurements. Consequently we have full heat dissipation between pulses, and the static sample temperature does not increase when we increase the pump fluence. Thus we are certain that no continuous heating of the sample occurs.

\begin{figure}
\centering
\includegraphics{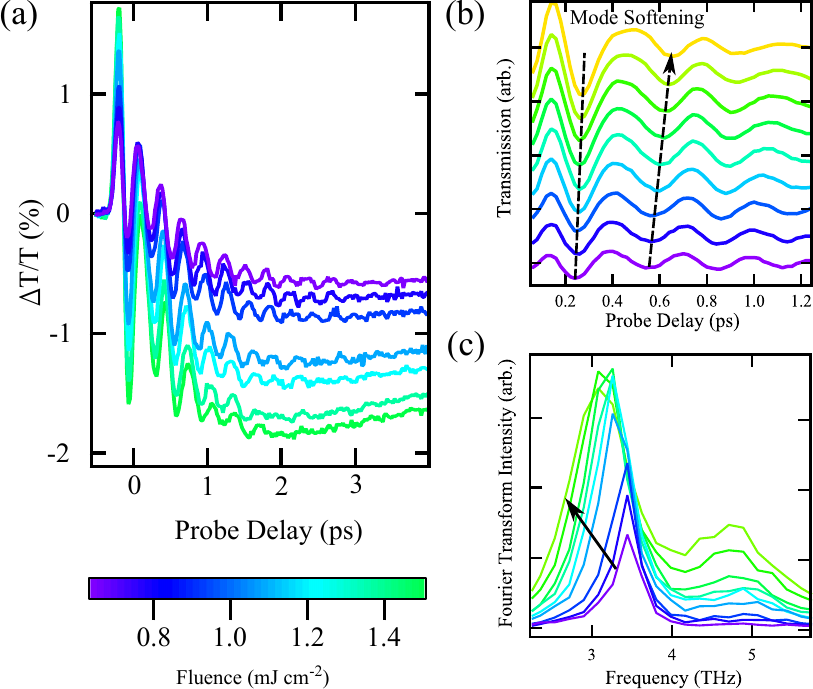}
\caption{(color online) Phonons in an amorphous sample exposed to prolonged periods of laser exposure without complete relaxation between pulses.  In subsequent measurements, the phonons ring longer (a). Also, phonon softening can clearly be seen with increasing fluence (b). The softening is clear in the Fourier transform as well, in addition to spectral weight changes (c). This indicates permanent sample change, and is consistent with Te segregation.\label{fig:phononSoftening}}
\end{figure}

We found that samples which had Raman spectra indicating Te segregation {\em did} show phonon softening with increasing pump fluence. To demonstrate this, we show in Figure~\ref{fig:phononSoftening} how the response of the amorphous phase changes after several hours of exposure to femtosecond laser pulses {\em without} inter-pluse relaxation. Upon subsequent reduction of pump energy, the overall changes in transmission closely match that of the pristine amorphous state shown in Figure~\ref{fig:RTgstAC}, but much longer lived phonons are seen. The observed frequency of the phonons clearly does depend on the pump power, unlike the data reported here. It is known that phonons in Te, like bismuth, are strongly dependent on the pump intensity~\cite{Zeiger1992}; since the observed phonons correspond to those of elemental Te~\cite{Amirtharaj1984, Xi2012}, we suspect that the laser induces permanent Te segregation without crystallizing the sample, and that isolated Te islands do not exhibit as strong damping as the localized phonons in GST. Similarly, we could also induce changes in the crystalline state phonons without amorphizing the sample on a macroscopic scale. Other groups have also reported that mode softening depends on sample preparation~\cite{Hernandez-Rueda2011a}. Thus careful characterization of the sample before and after the experiments is required. 

Finally we note that many experiments which show mode softening are on superlattice samples~\cite{Makino2011, Fons2015b}. It is known that the phase transition in GST can be controlled by strain, and that superlattices offer a way to control strain in the samples~\cite{Kalikka2016}. In superlattices, acoustic phonons become Raman active due to back-folding at the zone boundary. Although not present in the samples reported here, back-folded modes may be more strongly effected by pump-induced strain and show different dynamics than what we observe. As a result, we believe that coherent phonons in unstructured films cannot be used to drive amorphization; however superlattices may present different opportunities. 

\section{Conclusions}
In this work we have addressed the photoinduced changes in the dielectric function of both the amorphous and crystalline phases of GST. We find that the response at 800\,nm is sensitive to both structural and electronic effects. Although the measured transient reflectivity and transmission look qualitatively different in the two phases, conversion to the dielectric function reveals that the underlying dynamics are similar.

In both cases, the response is dominated by photo-excited carriers. In GSTa the carriers induce a Drude response at 800\,nm. Whereas, in GSTc, the photoexcited carriers depopulate resonant bonds, altering the optical matrix elements responsible for the dielectric response. Resonant bonding is then further destabalized by incoherent phonons, and GSTc does not recover until the excitation can be thermally dissipated.

In addition coherent acoustic and optical phonon modes also appear in the transient response. In the range of fluences measured here, we do not see any change in the frequencies of these oscillations, which suggests that the lattice potential and covalent bonding are relatively unaffected by photoexcitation at levels up to 50\% of the single-shot transformation fluence. We argue that the coherent optical phonons primarily modulate the resonant bonding-antibonding transition, and that, as these modes are localized near vacancies, they do not significantly perturb extended resonant bonding in the crystalline state.

Practically, the dielectric function changes in both the amorphous and crystalline states are of a similar amplitude. Thus the starting phase of choice will depend on whether an application requires large changes in transmission or reflection. However, the excited electrons persist longer in the crystalline state. As lattice heating drives the permanent phase transition, the crystalline phase may present better opportunities for extracting energy from the electronic system before it is transferred to the lattice, enabling larger transient changes to be induced without permanent structural modifications. 

\begin{acknowledgments}
We acknowledge fruitful discussions with Lutz Waldecker and Ralph Ernstorfer. We acknowledge financial support from the Spanish Ministry of Economy and Competitiveness through the `Severo Ochoa' Program for Centers of Excellence in R\&D (SEV-2015-0522) and Fundaci{\'o} Privada Cellex. TAM acknowledges funding through the ICFONEST+ program, funded by the Marie Curie COFUND project FP7-PEOPLE-2013-COFUND. VP acknowledges financial support from the `Fondo Europeo de Desarrollo Regional' (FEDER) through grant TEC2013-46168-R.  SW acknowledges financial support from the Ramon y Cajal program RYC-2013-14838 and Marie Curie Career Integration Grant PCIG12-GA-2013-618487. V.P. and S.W. acknowledge financial support from AGAUR 2014 SGR 1623.
\end{acknowledgments}

\end{document}